\documentclass[twocolumn,showpacs,amsmath,nofootinbib,amssymb,superscriptaddress]{revtex4}

\usepackage{graphicx,amsmath,amsfonts,amssymb,revsymb,dcolumn,epsfig,bm}

\begin{document}


\title{Future dynamics in $f(R)$ theories}

\author{D. M\"uller} 
\altaffiliation[On leave of absence from ]{Instituto de F\'{\i}sica, Universidade de Bras\'{\i}lia,
Bras\'ilia -- DF,  Brazil.}
\affiliation{CERCA/Department of Physics/ISO, Case Western Reserve University \\
Cleveland, OH 44106-7079, USA}
\author{V.C. de Andrade} 
\author{C. Maia} 
\affiliation{Instituto de F\'{\i}sica, Universidade de Bras\'{\i}lia  \\
70919-970 Bras\'ilia -- DF,  Brazil}  

\author{M.J. Rebou\c{c}as} 
\author{A.F.F. Teixeira}
\affiliation{Centro Brasileiro de Pesquisas F\'{\i}sicas,
Rua Dr.\ Xavier Sigaud 150 \\
22290-180 Rio de Janeiro -- RJ, Brazil}

\begin{abstract}
The $f(R)$ gravity theories provide an alternative way to explain the current cosmic
acceleration without invoking dark energy matter component.
However, the freedom in the choice of the functional forms of $f(R)$
gives rise to the problem of how to constrain and break the degeneracy among
these gravity theories on theoretical and/or observational grounds.
In this paper to proceed further with the investigation on the potentialities,
difficulties and limitations of $f(R)$ gravity,  we examine the question
as to whether the future dynamics can be used to break the degeneracy between
$f(R)$ gravity theories by investigating the future dynamics of spatially homogeneous
and isotropic dust flat models in two  $f(R)$
gravity theories, namely the well known  $f(R) = R + \alpha R^{n}$ gravity
and another by A. Aviles \emph{et al.}, whose motivation comes from
the cosmographic approach to $f(R)$ gravity. To this end we perform a
detailed numerical study of the future dynamic of these flat model in
these theories taking into account the recent constraints on the cosmological
parameters made by the Planck team. We show that besides being powerful for
discriminating between $f(R)$ gravity theories, the future dynamics technique
can also be used to determine the fate of the Universe in the framework of these
$f(R)$ gravity theories.
Moreover, there emerges from our numerical analysis that
if we do not invoke a dark energy component with equation-of-state parameter
$\omega < -1$ one still has  dust flat FLRW solution
with a big rip, if gravity deviates from general relativity via
$f(R) = R + \alpha R^n $. We also show that FLRW dust solutions  with $f''<0$
do not necessarily lead to singularity.

\end{abstract}


\maketitle

\section{Introduction} \label{Intro}

A wide range of cosmological observations coming from different sources,
including the supernovae type Ia (SNe Ia)~\cite{SNE}, the cosmic microwave background
radiation (CMBR)~\cite{CMB} and  baryon acoustic oscillation (BAO) surveys~\cite{BAO},
clearly indicate that the Universe is currently expanding with an accelerating rate.
A fair number of models and  frameworks have been proposed to account
for this observed accelerated expansion. These approaches can be roughly grouped in two
families. In the first, the so-called dark energy is invoked and the underlying
framework of general relativity (GR) is kept unchanged.
In this regard, the simplest way to account for  the accelerating expansion of the Universe
is through the introduction of a cosmological constant, $\Lambda$, into Einstein's field
equations. This is entirely consistent with the available observational data, but
it faces difficulties such as the order of magnitude of the cosmological constant and
its microphysical origin.
In the second family, modifications of Einstein's field equations are assumed
as an alternative for describing the accelerated expansion. This latter group
includes, for example, generalized theories of gravity based upon modifications of
the Einstein-Hilbert action by taking nonlinear functions, $f(R)$, of the Ricci
scalar $R$  or other curvature invariants (for reviews see Refs.~\cite{fr-reviews}).

The fact that $f(R)$ theories can potentially be used to explain the observed accelerating
expansion have given birth to a number of articles on these gravity theories, in which
several features of $f(R)$ gravity have been discussed~\cite{Some-fr-refs}, including
the stability conditions~\cite{Dolgov}, compatibility with solar-system tests~\cite{Chiba},
energy conditions~\cite{EC}, nonlocal causal structure~\cite{RS}, and observational
constraints from a diverse set of cosmological observations~\cite{Amarzguioui}.

However, although the freedom in the choice of the functional forms of $f(R)$
has motivated many different suggestions of $f(R)$ gravity theories, which account for
the accelerating expansion and are compatible with the solar-system tests,
it also gives rise to the problem of how to constrain and break the degeneracy among
these gravity theories on theoretical and/or observational grounds. In this regard,
observational constraints on some $f(R)$ gravity from a diverse set of
observations have been placed~\cite{Amarzguioui}, and tests of the cosmological
viability of some specific forms of $f(R)$ have been explored~ \cite{Amendola}.

A pertinent question that arises here is whether the future dynamics can be used
to break the degeneracy between  $f(R)$ gravity theories.
In this article, to proceed further with the investigation on the potentialities,
difficulties and limitations of $f(R)$ gravity,  we examine this question
by investigating the future dynamics of Friedmann-Lema\^{\i}tre-Robertson-Walker
(FLRW) dust flat model in two  $f(R)$ gravity theories,
namely the well known  $f(R) = R + \alpha R^{n}$ gravity, for which many results
are available in the literature~\cite{manyresults}, and another by A. Aviles
\emph{et al.}~\cite{ABCL} (ABCL gravity for short), whose motivation comes from
the cosmographic approach to $f(R)$ gravity~\cite{Cap,Star}. We  show that
besides being powerful for discriminating between $f(R)$ gravity theories, the
future dynamics technique can also be used to determine the fate of the Universe
in $f(R)$ gravity theory.


Until the discovery of the accelerating expansion virtually any textbook on
cosmology describes the  future dynamics of FLRW pressure-free dust models as follows.
It expands forever if it has an Euclidean or hyperbolic spatial geometry, and
expands and eventually recollapses if it has a spherical spatial geometry.
However, the discovery of the accelerating expansion made apparent that
these simple future forecasts  had to be modified,
since the negative-pressure dark energy component, invoked to account
for the acceleration,  plays a crucial role in the evolution of the Universe.
Indeed, the dark energy (DE) is usually described by the equation-of-state parameter
$\omega= p/\rho$, which is the ratio of the DE pressure to its density. A value
$\omega < -1/3$ is required for cosmic acceleration. When $ -1 < \omega < -1/3$
the DE density decreases with the scale factor $a(t)$. However, if $\omega <-1$
the dark energy  density becomes infinite in a finite-time, driving therefore the
Universe to a future finite-time singularity, called big rip~\cite{Caldwell-Starobinsky}.
Afterwards it was determined that this is not the only possible doomsday of a dark
energy dominated universe. It may, for example, come to an end in a sudden
singularity~\cite{Sudden}, a big freeze doomsday~\cite{Bigfreeze},
or a little rip~\cite{Litterip}.
In this paper we also determine that even if we do not invoke a dark energy
component with $\omega < -1$ one still has pressure-free dust solution
with a big rip, if gravity deviates from general relativity via
$f(R) = R + \alpha R^n $.
Finally, by using the future dynamics scheme of this paper, we further present
an example of FLRW dust solution  in which the ghost-like regimes ($f''<0) $
do not necessarily lead to singularity%
\footnote{We note that future dynamics in $f(R)$ gravity was also considered
in Ref.~\cite{Nojiri-Odintsov}. Their  approaches are different from ours
since they have considered perfect fluid matter source with
$\rho, p \;\,\mbox{and}\;\, \omega=p/\rho$, while we consider simply a pressure-free
dust. Furthermore, our analysis is numerical while theirs is not.}.

Our paper is organized as follows. In Sec.~\ref{Basics} we give a brief review
of $f(R)$ gravity theories, derive the field equations for the flat
FLRW metric with dust matter content, state the initial conditions for the
dynamical evolutions, and
present the future dynamics in the context of the general relativity theory.
In Sec.~\ref{ABCL-sec} we introduce the ABCL~\cite{ABCL}
Lagrangian, develop the necessary technique for solving the dynamical equations,
and derive our numerical results regarding the  ABCL $f(R)$ gravity.
In Sec.~\ref{rn} we use our method to study the polynomial Lagrangian $f(R)$
gravity and make a comparative analysis of these theories.
Final remarks and main conclusions are presented in Sec.~\ref{conc}.

\section{Prerequisites} \label{Basics}

In this section we briefly review the $f(R)$ gravity, derive the field equations
for the flat FLRW metric with dust matter content, state the initial conditions
for all numerical analyses of this paper, and present the future dynamics
for the particular Einstein's gravity theory for later comparison with
the dynamics in other gravity theories.

\subsection{$\mathbf{f(R)}$ gravity and field equations}

We begin by recalling that the action that defines an $f(R)$ gravity theory
can be written as

\begin{equation}  \label{action}
 S=\frac{1}{2\kappa^2}\int d^4x\sqrt{-g}\,f(R) + S_m \,,
\end{equation}
where $\kappa^2\equiv 8\pi G/c^4$, $g$ is the determinant of the metric
$g_{ab}$,  $f(R)$ is a function of the Ricci scalar $R$, and $S_m$ the
standard action for the matter fields. Varying this action with
respect to the metric we obtain the field equations
\begin{equation} 
f'R_{ab} - \frac{f}{2}g_{ab} - \left(\nabla_{a}\nabla_{b}-
g_{ab}\,\Box \,\right)f' = \frac{8\pi G}{c^{4}}T_{ab}\,, \label{eq campo}
\end{equation}
where here and in what follows primes denotes differentiation with respect to $R$ 
and $\Box \equiv g^{ab}\,\nabla_{a}\nabla_{b}\,$.
Clearly, for $f(R)=R+ \Lambda$ these field equations reduce to the Einstein equations with
the cosmological constant, $\Lambda$, term.

Two important constraints, often used to simplify the calculations, come from
the fact that the covariant divergence of both sides of Eq.~\eqref{eq campo},
is null. This  implies that the $0i$ and $00$ components of the field
equations give rise to the following constraints
\begin{eqnarray}
E_{0i}&=&f^{\prime}R_{0i}-\frac{f}{2}g_{0i}-(\nabla_{0}\nabla_{i}-g_{0i}
\square)f^{\prime}=0\,, \label{1st-constraint} \\
E_{00} &=&f^{\prime}R_{00}-\frac{f}{2}g_{00}-(\nabla_{0}\nabla_{0}
-g_{00}\square)f^{\prime} \nonumber\\
& & - \,\frac{8\pi G}{c^{4}}\,\,T_{00}=0\,, \label{2nd-constraint} 
\end{eqnarray}
Clearly, Eq.~\eqref{1st-constraint} is identically satisfied, while the
constraint given by Eq.~\eqref{2nd-constraint} must be fulfilled throughout
time evolution. We shall use this fact as a way of checking the accuracy
of the numerical integration of the remaining equations of the dynamical
system in the numerical analyses of the following sections.

In this work we focus on the flat Friedmann-Lema\^{\i}tre-Robertson-Walker (FLRW) metric,
\begin{equation}
ds^{2}=-c^2dt^{2}+a(t)^{2}\left(dx^{2}+dy^{2}+dz^{2}\right)\,, \label{FLRW-metric}
\end{equation}
which is supported by the recent observations~\cite{CMB}, and is consistent with
the standard inflationary models.
Thus, the non vanishing component of the Ricci tensor and the Ricci scalar
can be written in the form
\begin{eqnarray}
&&R^0_0=\frac{3}{c^2}\left( \dot{H}+H^2 \right)\,,\label{R00} \\
&&R_1^1=R_2^2=R_3^3=\frac{1}{c^2}\left(\dot{H}+3H^2\right)\,, \label{Rij} \\
&&R=\frac{6}{c^2}\left( \dot{H}+2H^2\right)\,, \label{R}
\end{eqnarray}
where $H=\dot{a}/a$ and the over-dot denotes derivative with respect to the time $t$.

Since we are interested in dust of density $\rho$ with zero pressure ($p=0$) then
we have
\begin{equation} \label{rho-eq}
 \dot{\rho}=-3H\rho  \quad \; \mbox{and} \; \quad \rho=\frac{\rho_0}{a^3}\,,
\end{equation}
where here and in what follows the subscript zero denotes present-day values of
the cosmological parameters.

Taking into account equations~\eqref{R00}--\eqref{rho-eq} the field equations
\eqref{eq campo} reduce to
\begin{equation}
-3\left(\dot{H} +H^2\right)f^\prime
+\frac{fc^2}{2}+3Hf^{\prime\prime}\dot{R}=\frac{8\pi G\rho}{c^2}
\label{eqfmod0} \,,   \\
\end{equation}
\begin{equation}
f^\prime\left(\dot{H}+3H^2\right)-
\frac{fc^2}{2}-2Hf^{\prime\prime}\dot{R}-f^{\prime\prime}\ddot{R}   
 -\dot{R}^2f^{\prime\prime\prime}=0\,.  
\label{eqfmod1}
\end{equation}
One can easily show that Eq.~\eqref{eqfmod0} is nothing but the constraint
Eq.~\eqref{2nd-constraint}, which for the dust flat FLRW models  takes
the form
\begin{equation}
E_{00}=-3\left(\dot{H} +H^2\right)f^\prime
+\frac{fc^2}{2}+3Hf^{\prime\prime}\dot{R}-\frac{8\pi G\rho}{c^2}=0\,,
\label{Energia}
\end{equation}
which is in a suitable form for checking the accuracy of the numerical
integration of the dynamical Eq.~\eqref{eqfmod1} for the  
$f(R)$ gravity theories we are concerned with in this paper.

\begin{figure*}[thb!] 
\begin{center}
\includegraphics[width=6.5cm,height=4.0cm]{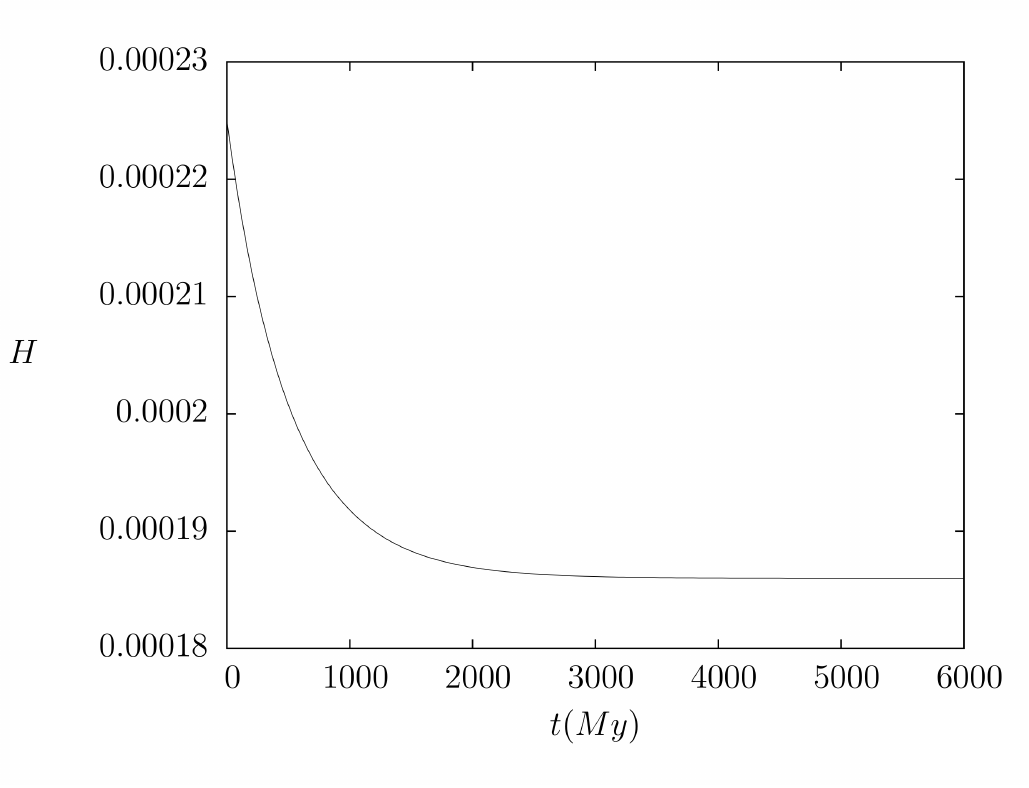} 
\hspace{8mm}
\includegraphics[width=6.5cm,height=4.0cm]{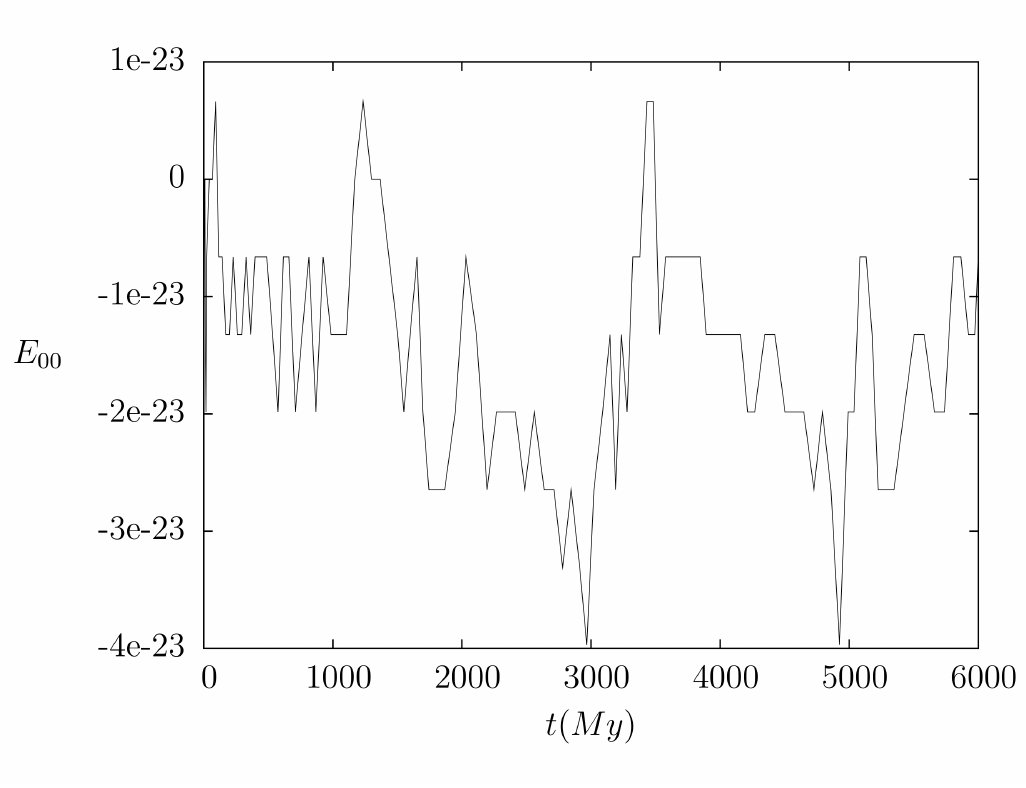} {f2.ps}
\caption{\textbf{Left panel}: Time evolution of $H$ for spatially flat FLRW dust model in the
general relativity. \textbf{Right panel}: The behavior of the constraint \eqref{Energia}
making apparent the high level of accuracy in the numerical integration has been obtained.
The current values of the cosmological parameters as given by Planck team~\cite{CMB} were taken as
initial conditions in the numerical integration. \label{Fig1}}
\end{center}
\end{figure*}

\subsection{Initial conditions} \label{Ini-cond}

\begin{table}[tbh]
\begin{center}
\begin{tabular}{cc} 
\hline \hline
Cosmological parameters \ \  & \ \ \ \ \ \ \ \ Values \\
\hline
$H_0$                        &  $(2.25\pm 0.05)\times 10^{-4}1/T$~\cite{CMB} \\
$\Omega_\Lambda$             &  $0.686\pm 0.020$~\cite{CMB} \\
$\Omega_m$                   &  $0.314\pm0.020$~\cite{CMB} \\
$q_0$                        &  $-0.81\pm0.14$ \cite{rapetti}\\
$j_0$                        &  $2.16_{-0.75}^{+0.81}$~\cite{rapetti} \\
$\rho$                       &  $(4.0\pm 0.5)\times10^{10}\frac{M_{\odot}}{Mpc^{3}}$ \\
\hline \hline
\end{tabular}
\end{center}
\caption{Values of the cosmological parameters used as initial conditions in
the numerical integrations.
As a suitable choice of units, the length is the $Mpc$, the time unit $T=3.26158\times10^{6}y$
and the mass unit is chosen as $M_\odot$, so that in these units the speed of light $c=1$,
and Newton's constant is $G=4.7863\times10^{-20}\frac{Mpc^{3}}{M_{\odot}T}$.
The values of the cosmological parameters, deceleration and jerk parameters,  are taken
from the Planck results~\cite{CMB} from  Ref.~\cite{dabrowski}.  \label{Table1} }
\end{table}

To study the future dynamics for the spatially flat FLRW dust models one needs
to choose initial conditions for the numerical integration. In this work we use
the numerical values of the cosmological parameters reported by the  Planck
Collaboration team~\cite{CMB} along with the values of the cosmographic
parameters given in Ref.~\cite{dabrowski}. In Table~\ref{Table1} we collect
together the values of the cosmological parameters we shall employ in
our numerical analyses. Table~\ref{Table1} also contains details of the
units and convention we have adopted in this paper.

To investigate the future dynamics in the following sections, we recall that for the flat
FLRW models the dimensionless deceleration $(q)$ and jerk $(j)$ parameters~\cite{dabrowski}
given  in Table~\ref{Table1} are such that the relations
\begin{eqnarray}
 &  & q_{0}=-\frac{1}{H_{0}^{2}}\left(\dot{H}_0+H_0^2\right)\,, \label{q0} \\
 &  & j_{0}=\frac{1}{H_{0}^{3}}\left(\ddot{H}_0+3H_0\dot{H}_0+H_0^3\right) \label{j0} \,.
\end{eqnarray}
hold.

\subsection{Dynamics in General Relativity}

For a later comparison with the dynamics in other gravity theories,
we  briefly present here the analysis for the spatially flat FLRW dust
model in the Einstein theory with cosmological constant $\Lambda$, that
is for $f(R)=R+\Lambda$. In this case the field equations \eqref{Energia}
and Eq.~\eqref{eqfmod1} reduce, respectively,  to   
\begin{eqnarray}
E_{00} = 3H^2 + \frac{\Lambda c^2}{2} - \frac{8\pi G}{c^2}\rho &= &0\,, \\
-2\dot{H} - 3H^2 -\frac{\Lambda c^2}{2}& = &0\,.
\end{eqnarray}

The left panel in Figure~\ref{Fig1} shows the evolution of $H$ for spatially flat
FLRW dust model in the Einstein theory, where initial conditions given in
Table~\ref{Table1}  were employed in the numerical calculations. This panel also
shows a de Sitter asymptotic behavior for $H$ for a non vanishing cosmological
constant $\Lambda$. The right panel shows the constraint equation~\eqref{Energia},
providing an assess of the reliability of the numerical integration.

\section{Dynamics in ABCL gravity \label{ABCL-sec}}

In this section we shall study the future dynamic of the spatially flat FLRW dust
model in the $f(R)$ gravity recently suggested by A. Aviles \emph{et al.}~\cite{ABCL},
referred to in this paper as ABCL gravity theory. This gravity theory has been obtained
through an optimal Monte-Carlo fitting of cosmographic results and is given by
\begin{widetext}
\begin{eqnarray}
&&f(R)=\frac{1}{2(a+b+c)e\pi R_0^2}\left\{
\Lambda R_0^2\left[ 2a\pi e^{R/R_0}+e\left( 6b+(a+2c)\pi
+8b\arctan \left(\frac{R}{R_0}\right) \right) \right]
\right. \nonumber\\
&&\left. +eR\left[ 2R_0 \left(\,(a+b+c)\pi R_0-4b\Lambda\,\right)+(2b-a\pi )\Lambda R\right]
-2ce\pi \Lambda(R-R_0)^2\sin \left( \frac{2\pi R}{R_0}\right)\right\},
\label{t_abcl}
\end{eqnarray}
\end{widetext}
where $a$, $b$ and $c$ are free parameters, and $R_0$ is the present-day value of
the Ricci scalar. Regardless of the values of these parameters for this theory one
has
\begin{eqnarray}
f(R_0)&=&R_0+\Lambda\,, \label{f_0}\\
f^\prime_0&=&1\,,\label{f_1}\\
f^{\prime\prime}_0&=&0\,, \label{f_2}
\end{eqnarray}
where $f^{\prime}_{0} \equiv (\partial f/\partial R)_{R = R_{0}}$ and similar
notation is used for higher order derivatives. In addition to the
constraints~\eqref{f_0} and~\eqref{f_1}, which are required to ensure that
both Einstein's theory and Newton's constant are recovered in the lowest order,
we shall take into account that $f^{\prime\prime}\geq 0$, which is a condition
to avoid the presence of ghosts~\cite{dolsol}. We also note that
the conditions \eqref{f_0}--\eqref{f_2} also insure that
\eqref{eqfmod0} and \eqref{eqfmod1} reduce to the Friedmann equations
in the lower order.

In order to study the dynamics we will need

\begin{eqnarray}
&&f^{\prime\prime\prime}_0=\Lambda\frac{2b+\pi(a-12c\pi)}{(a+b+c)\pi
R_0^3}\nonumber,\\
&&f_0^{iv}=\frac{a\Lambda}{(a+b+c)R^4_0}\nonumber,\\
&&f_0^{v}={\frac {\Lambda a\pi -12b\Lambda+160\pi ^{4}c\Lambda}{(a +
 b+ c)\pi R_0^5}}\nonumber,\\
&&f_0^{vi}={\frac {\Lambda a\pi +60b\Lambda}{(a + b+ c)\pi R_0^6}}\nonumber,\\
&&f_0^{vii}={\frac {\Lambda a\pi -180b\Lambda-1344\pi^{6}c\Lambda}{(a +
 b+ c)\pi R_0^7}},\nonumber\\
&&f_0^{viii}=\frac{\Lambda a}{(a+b+c)R_0^8}.
\end{eqnarray}

Before proceeding to the numerical analysis for this gravity theory, we
note that due care ought to be taken in using the initial conditions
since $f^{\prime\prime}=0$ at $t=t_0$. Note that in equations~\eqref{eqfmod0} and
\eqref{eqfmod1} the higher derivatives of the scalar factor are multiplied
by $f^{\prime\prime}$.
When $f^{\prime\prime}\neq 0$, we have a set of differential equations describing
the dynamics. When $f^{\prime\prime}=0$ locally, at $t=t_0$ we have an entirely
different set of differential equations obtained by~\eqref{eqfmod0} and~\eqref{eqfmod1}.
In what follows, to deal with this difficulty we first assume that the solution
possesses a Taylor expansion about $t=t_0$ up to second order. Then we substitute
this expansion into the field equations~\eqref{eqfmod0} and~\eqref{eqfmod1}
in an order by order manner,  which results in a perturbative solution up to
some order. Second, instead of assuming the initial condition exactly at $t=t_0$,
the initial condition is taken at $t=t_0+\epsilon$ ($\epsilon^2 \lll 1$) through
this perturbative scheme, for which now $f^{\prime\prime}(t_0+\epsilon)\neq 0$.

To carry out the outlined perturbative procedure, it is required to distinguish
two different regimes in establishing the initial condition ($f'=(t_0+\epsilon)\neq 0$),
namely one when at $t = t_0$, $f^{\prime\prime\prime} \neq 0$,  and another when
$t = t_0$, $ f^{\prime\prime\prime}=0$. In the following we shall treat separately
these two cases.

\subsection{The case $\mathbf{f_0^{\prime\prime\prime}\neq 0}$ \label{f0neq}}

Since for the ABCL gravity theory $f^{\prime\prime}(t_0)=0$, in order to find a
suitable form for the field equations within a perturbative scheme, we first
assume that the solution
can be expanded in a Taylor series about $t=t_0$ up to second order. Thus,
to obtain the terms of \eqref{eqfmod0} and \eqref{eqfmod1} we have

\begin{eqnarray}
&&H=H_0+\dot{H}_0(t-t_0)+\frac{1}{2}\ddot{H}_0(t-t_0)^2,\nonumber\\
&&H^2=H_0^2+2H_0\dot{H}_0(t-t_0)+ ( \dot{H}_0^2+ H_0\ddot{H}_0)(t-t_0)^2,\nonumber\\
&&\dot{H}=\dot{H}_0+\ddot{H}_0(t-t_0)+\frac{1}{2}\dddot{H}_0(t-t_0)^2,
\nonumber\\
&&\dot{R}^2=\dot{R}_0^2+2\dot{R}_0\ddot{R}_0(t-t_0)+ (\ddot{R}
_0^2+\dot{R}_0\dddot{R}_0 )(t-t_0)^2,\nonumber\\
&&\dot{R}=\dot{R}_0+\ddot{R}_0(t-t_0)+\frac{1}{2}\dddot{R}_0(t-t_0)^2,
\nonumber\\
&&\ddot{R}=\ddot{R}_0+\dddot{R}_0(t-t_0)+\frac{1}{2}\ddddot{R}_0(t-t_0)^2,
\nonumber\\
&&\rho=\rho_0-3H_0\rho_0(t-t_0)+\frac{1}{2}(9H_0^2-3\dot{H}_0)\rho_0(t-t_0)^2
,\nonumber\\
&&f=R_0+\Lambda+\dot{R}_0(t-t_0)+\frac{\ddot{R}_0}{2}(t-t_0)^2,\nonumber\\
&&f^\prime=1+\frac{1}{2}f^{\prime\prime\prime}_0\dot{R}_0^2(t-t_0)^2,
\nonumber\\
&&f^{\prime\prime}=f^{\prime\prime\prime}_0\dot{R}_0(t-t_0)+\frac{1}{2}
(f^{iv}_0\dot{R}_0^2+f^{\prime\prime\prime}_0\ddot{R}_0)(t-t_0)^2,\nonumber\\
&&f^{\prime\prime\prime}=f^{\prime\prime\prime}_0+f_0^{iv}\dot{R}_0(t-t_0)+
\frac{1}{2}(f^{v}_0\dot{R}_0^2+f_0^{iv}\ddot{R}_0)(t-t_0)^2. \nonumber \label{expansao}
\end{eqnarray}
%
\begin{figure*}[thb!] 
\begin{center}
\includegraphics[width=6.5cm,height=4.0cm]{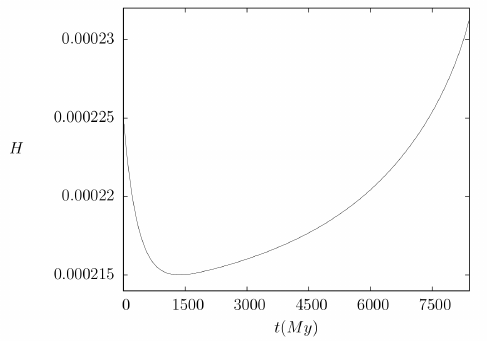}  
\hspace{8mm}
\includegraphics[width=6.5cm,height=4.0cm]{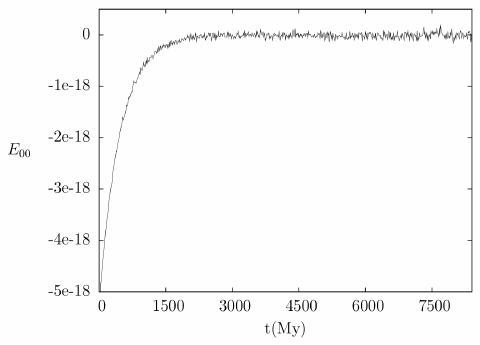}  
\caption{\textbf{Left panel}: Time evolution of $H$ for spatially flat FLRW dust model in the
ABCL gravity theory with the initial conditions specified by eqs.~\eqref{ci1}-\eqref{ci3}.
The values of the free parameters were taken to be $a=107.9$, $b=-148.0$ and $c=40.0$,
and the values of the cosmological parameters are the best fit values collected together
in Table~\ref{Table1} as given by Planck team~\cite{CMB} were taken as
initial conditions in the numerical integration.
\textbf{Right panel}: The behavior of the constraint $E_{00}$ as given by~\eqref{Energia},
making apparent the degree of confidence  of the numerical calculations. No singularities were found
throughout the numerical evolution. \label{Fig2} }
\end{center}
\end{figure*}

Then we substitute these terms into the field equations~\eqref{eqfmod0}
and~\eqref{eqfmod1} in an order by order mode to have
\begin{itemize}
\item Zero order in $(t-t_0)$. Equations~\eqref{eqfmod0} and~\eqref{eqfmod1}
give
\begin{eqnarray}
&& 3(H_0)^2+\frac{\Lambda c^2}{2}=\frac{8\pi G\rho_0}{c^2}\,, \label{0st-a} \\
&& -2\dot{H}_0-3(H_0)^2-\frac{\Lambda c^2}{2}-f_0^{\prime\prime\prime}\dot{R}_0=0\,. \label{0st-b}
\end{eqnarray}
\item First order in $(t-t_0)$. Equations~\eqref{eqfmod0}~ and~\eqref{eqfmod1}
yield
\begin{eqnarray}
&&-3(\ddot{H}_0+2H_0\dot{H}_0)+\frac{c^2}{2}\dot{R}_0
+3 H_0f_0^{\prime\prime\prime} \dot{R}_0^2 \nonumber\\
&&=\frac{8\pi G}{c^2}(-3H_0\rho_0) \,, \label{1st-a} \\
&& (\ddot{H}_0+6H_0\dot{H}_0)-\frac{c^2}{2}\dot{R}_0
-f_0^{\prime\prime\prime}\dot{R}_0  (2H_0\dot{R}_0 +3\ddot{R_0}) \nonumber \\
&&-f_0^{iv}(\dot{R}_0)^3=0\,.\label{1st-b}
\end{eqnarray}
\end{itemize}
{}From \eqref{0st-a}--\eqref{1st-b} one has
\begin{eqnarray}
&&H_0^2=\frac{1}{3}\left(\frac{8\pi G\rho_0}{c^2} -\frac{\Lambda c^2}{2}\right)\,,
\label{H0dH0-a} \\ 
&&\ddot{H}_0=\frac{c^2}{6}\left(\frac{-2\dot{H}_0-3H_0^2-(\Lambda c^2)/2}{f_0^{'''}}\right)^{1/2} \nonumber \\ && -4H_0\dot{H}_0\,, \label{H0dH0-b}\\ 
&&\dddot{H}_0=-4 \dot{H}_0^2-4H_0\ddot{H}_0 \nonumber \\
&&-\frac{c^2}{6}\left(\frac{2\ddot{H}_0+6H_0\dot{H}_0
+2H_0f_0^{\prime\prime\prime}\dot{R}_0^2+f_0^{iv}\dot{R}_0^3}{3f_0^{\prime\prime\prime}\dot{R}_0}\right)\!, \label{H0dH0-c}
\end{eqnarray}
where from equation~\eqref{R} we have
\begin{equation} \label{dotR}
\dot{R}_0=6(\ddot{H}_0+4H_0\dot{H}_0)/c^2\,.
\end{equation}

As mentioned, if $f^{\prime\prime}\neq 0$ the initial condition
follows directly  from eqs.~\eqref{eqfmod0} and~\eqref{eqfmod1}.
Otherwise, if $f^{\prime\prime}_0=0$ at $t=t_0$, then the initial condition
is chosen for $t$ near $t_0$, so that we can be sure that
$f^{\prime\prime}(t)\neq 0$ and the dynamics is directly described by
eqs.~\eqref{eqfmod0} and~\eqref{eqfmod1}. In this case, instead of
taking $H_0$, $\dot{H}_0$ and $\ddot{H}_0$ the initial conditions
are chosen at $t-t_0=\epsilon$
\begin{eqnarray}
&&H=H_0+\dot{H_0}(t-t_0)+\frac{1}{2}\ddot{H}_0(t-t_0)^2,\label{ci1}\\
&&\dot{H}=\dot{H}_0+\ddot{H}_0(t-t_0)+\frac{1}{2}\dddot{H}_0(t-t_0)^2,\\
&&\ddot{H}=\ddot{H}_0+\dddot{H}_0(t-t_0), \label{ci3}
\end{eqnarray}
where $\dot{H}_0$, $\ddot{H}_0$, $\dddot{H}_0$ satisfy the relations
\eqref{H0dH0-a}--\eqref{H0dH0-c}.
We  note that according to eqs.~\eqref{H0dH0-b} the initial
value $\dot{H}_0$ ought to obey the constraint
\begin{equation}
\frac{-2\dot{H}_0-3H_0^2-(\Lambda c^2)/2}{f_0^{\prime\prime\prime}} <0.
\end{equation}

The left panel of Figure~\ref{Fig2} shows a representative numerical future
dynamics of spatially flat FLRW dust model in the ABCL gravity theory with
$f_0^{\prime\prime\prime} \neq 0$. In this case, as indicated by $H(t)$,
the universe would present a noteworthy expanding phase after an initial
future decelerating period. The right panel in this figure shows
the constraint  $E_{00}=0$, given in eq.~\eqref{Energia}, fluctuates
randomly and increases but is always smaller than $2\times 10^{-17}$,
which is a strong indication of the correctness of the numerical solution.

Regarding the choice of the $a$, $b$ and $c$ used to find the numerical future
dynamics solution, it is important to point out that the independence of equations
\eqref{f_0}--\eqref{f_2} allows some freedom in their choice.
Although a more general phase space analysis, with its associated attractors, would
be necessary to determine every possible future dynamics solution, here we have restricted
our analysis to a set of values which are consistent with the present-day constraints
on the cosmological parameters. This choice of values was also
motivated by a similar procedure used in the study of the polynomial $f(R)$ gravity,
which we study in details in the Section~\ref{rn}.

\subsection{The case $\mathbf{f_0^{\prime\prime\prime}=0}$\label{f'''=0}}

\begin{figure*}[thb!] 
\begin{center}
\includegraphics[width=6.5cm,height=4.0cm]{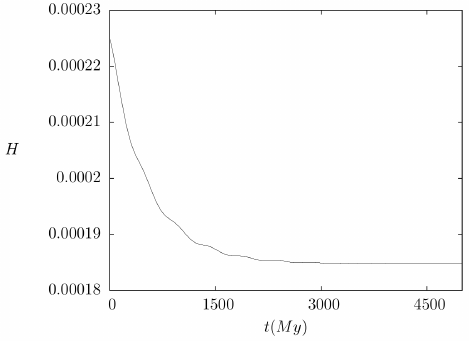}  
\hspace{8mm}
\includegraphics[width=6.5cm,height=4.0cm]{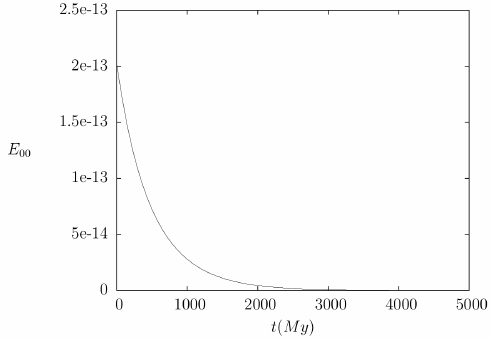}  
\caption{  \textbf{Left panel}: Time evolution of $H$ for spatially flat FLRW dust model in the
ABCL gravity theory. The values of the free parameters were taken to be $a=157.0$, $b=-258.5$ 
and $c=-0.2$, which fulfills the constraint equation~\eqref{cond_a}.  
The cosmological parameters were taken to be the best fit values collected in Table~\ref{Table1}
as given by Planck team~\cite{CMB}.
The dynamic evolution of the flat FLRW spacetime tends asymptotically to Sitter space.
\textbf{Right panel}: The behavior of the constraint $E_{00}$ as given
by~\eqref{Energia}, which is smaller than $10^{-13}$ for all time $t$, making apparent the
precision of the numerical calculation throughout evolution of the model. \label{Fig3} }
\end{center}
\end{figure*}

The condition $f_0'''=0$  can be achieved in the ABCL $f(R)$ gravity
provided that the constraint
\begin{equation} \label{cond_a}
a = 12 \pi c - \frac{2b}{\pi}
\end{equation}
holds.
Now, similarly to the previous section we assume that the solution
can be expanded in a Taylor series  $t=t_0$ up to second order.
Then we substitute Taylor series terms into the field equations~\eqref{eqfmod0}
and~\eqref{eqfmod1} in an order by order manner. This gives the following:
\begin{itemize}
\item
Zero order in $(t-t_0)$. Equations~\eqref{eqfmod0} and~\eqref{eqfmod1}
give
\begin{eqnarray}
&&3(H_0)^2+\frac{\Lambda c^2}{2}=\frac{8\pi G\rho_0}{c^2}, \label{Z-order-a}\\
&&-2\dot{H}_0=\frac{8\pi G\rho_0}{c^2} \label{Z-order-b}.
\end{eqnarray}
\item
First order in $(t-t_0)$. Equations~\eqref{eqfmod0}~ and~\eqref{eqfmod1}
yield
\begin{eqnarray}
&&-2\dot{H}_0=\frac{8\pi G\rho_0}{c^2}, \label{F-order-a}\\
&&f_0^{iv}\dot{R}_0^3+\frac{\dot{R}_0c^2}{3}-2H_0\dot{H}_0=0.\label{3grau}
\end{eqnarray}
\item
Second order in $(t-t_0)$. Equations~\eqref{eqfmod0}~ and~\eqref{eqfmod1}
furnish
\begin{equation}
6\dot{H}_0+6H_0\ddot{H}_0+3H_0f_0^{iv}\dot{R}_0^3=
\frac{8\pi G\rho_0}{c^2}\left(9H_0^2-3\dot{H}_0\right), \label{S-order-a}
\end{equation}
\begin{equation}
-2\dddot{H}_0-6\dot{H}_0^2-6H_0\ddot{H}_0-6f_0^{iv}\dot{R}_0^2\ddot{R}_0 -2H_{0}f_0^{iv}\dot{R}^{3} - f_0^v\dot{R}_0^4=0,  \label{S-order-b}
\end{equation} \end{itemize}

Now, since $\dot{R}$ is given by Eq.~\eqref{dotR} it is clear from eq.~\eqref{3grau}
that $\ddot{H}_0$  is given  by a third order algebraic equation. Thus, from the
above equations~\eqref{Z-order-a}--\eqref{S-order-b} one has
\begin{eqnarray}
&&H_0=\sqrt{\frac{8\pi G\rho_0}{3c^2}-\frac{\Lambda c^2}{6}},\nonumber\\
&&\dot{H}_0=-\frac{4\pi G\rho_0}{c^2},\nonumber 
\end{eqnarray}
and
\begin{equation}
\ddot{H}_0=-4H_{0}\dot{H}_{0}+\frac{c^{2}}{6}\left[\frac{1}{6f_{0}^{iv}}
  \sqrt[3]{\Delta\,\left(f_{0}^{iv}\right)^{2}}
  -\frac{2c^{2}}{3 \sqrt[3]{\Delta\,\left(f_{0}^{iv}\right)^{2}}} \right], \label{H0dH02}
\end{equation}
where
\begin{equation}
\Delta = 216H_{0}\dot{H}_{0}+12\sqrt{3}\,\left[\frac{4\left(c^{2}/3\right)^{3}
 +108\left(H_{0}\dot{H}_{0}\right)^{2}f_{0}^{iv}}{f_{0}^{iv}}\right]^{1/2}.
\end{equation}

We display in Figure~\ref{Fig3} another representative (a suitable choice of
$a$ $b$ $c$) numerical solution, but now for the case $f_0^{\prime\prime} = 0$.
The left panel of the figure shows the dynamic evolution of this dust flat FLRW
model tends asymptotically to de Sitter space. Its asymptote 
approaches the de Sitter vacuum solution obtained by analytically solving Eq.~\eqref{eq campo}
with no matter content, which returns a value $H=1.8476\times 10^{-4}$, i.e. the asymptote
shown in Figure~\ref{Fig3}. By carrying out a stability analysis having this solution
as background, it is possible to show this is
an attractor solution with $f^{\prime\prime}>0$. The right panel of Figure~\ref{Fig3}
makes clear that the constraint on $E_{00}$ [Eq.~\eqref{Energia}], is smaller than
$2\times10^{-13}$ during the evolution of the FLRW flat model. This makes apparent
the accuracy of the numerical calculations performed in the study of the future dynamics
in this case.

\section{Dynamics in  $\mathbf{ f(R) = R + \alpha R^{n} } $ } 
\label{rn}

In this section we apply the same numerical scheme used in the previous section
to study the well known $f(R)$ gravity theory
\begin{equation}
f(R)=R+\alpha R^{n}. \label{f(R)}
\end{equation}

The above $f(R)$ has the special feature of presenting equivalent results
to the $\Lambda$CDM model, when applied to a FLRW setting within a
range of choices for the constants $\alpha$ and $n$. In this sense
no observationally relevant prediction would distinguish these
cases as established in \cite{amendola-2}.

As referred to in the end of Sec.~\ref{f0neq}, a straightforward
calculation to guide the choice of parameters $\alpha$ and $n$ can be
achieved by using the field equations~\eqref{eqfmod0} along with
equations~\eqref{f(R)}, \eqref{q0} and~\eqref{j0}. Indeed, these
equations allow to write $\alpha$ as
\begin{widetext}
\begin{equation}
\alpha=\frac{-2c^{2n-2}\,\left( 3\,H_0^{2}-8\pi G\rho_0/c^2\, \right)
(1-q_0)^{2-2n}
}{ \left( 6\,H_0^{2}
 \right) ^{n} \left( 2\,n-2\,q_0+1+2\,nq_0-
{n}^{2}q_0-2\,{n}^{2}-nj_0+{n}^{2}j_0-n{q_0}^{2}+{q_0}^{2}\right) }\,,
\label{rest.alfa}
\end{equation}
\end{widetext}
Where observed values of parameters $H_0$, $q_0$,  $j_0$ and $\rho_0$ and
their uncertainties are given in Table~\ref{Table1}.

\begin{figure*}[] 
\begin{center}
\includegraphics[width=6.5cm,height=4.0cm]{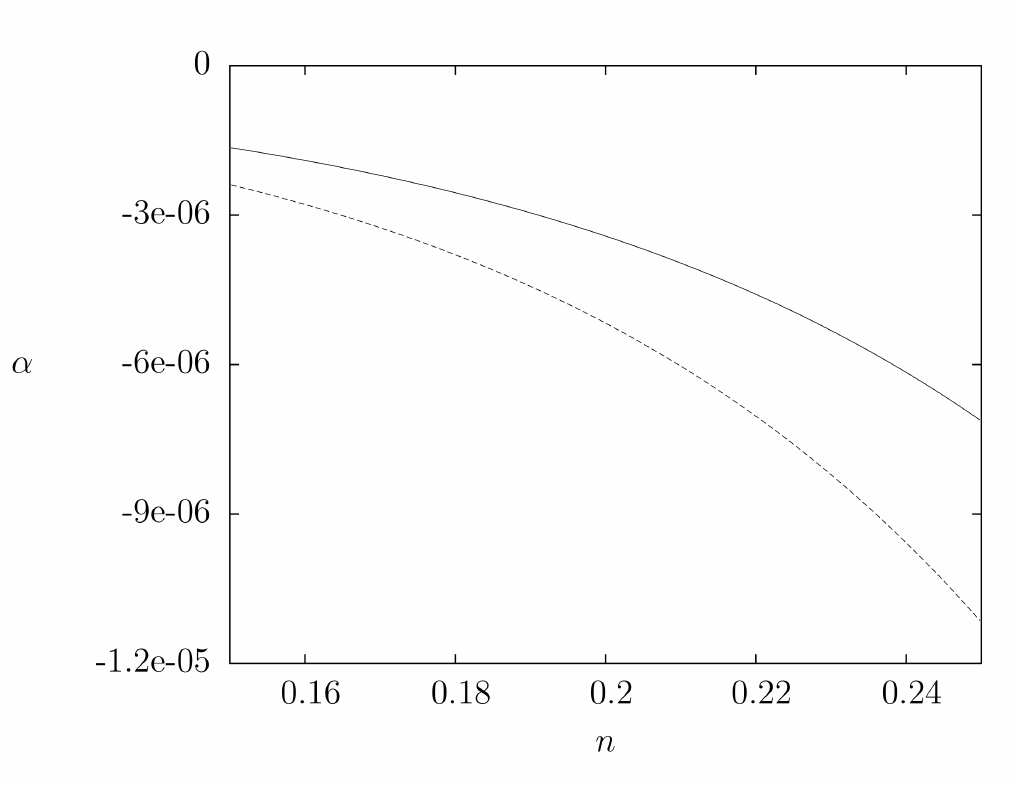}
\hspace{8mm}
\includegraphics[width=6.5cm,height=4.0cm]{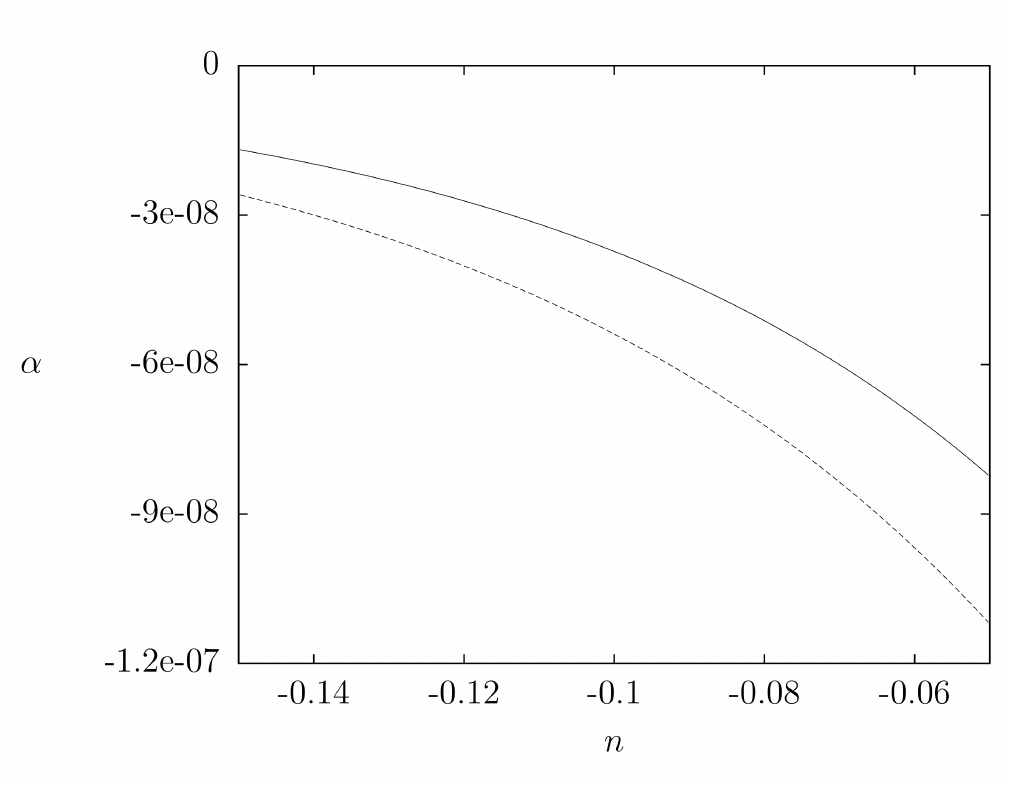}
\caption{\textbf{Left panel}: Bounds on the values of $\alpha$ for different values of $n$
derived from Eqs.~\eqref{rest.alfa} and~\eqref{dalpha} taking into account the values of
the cosmological parameters and associated uncertainties collected in Table~\ref{Table1}.
The depicted positive values of $n$ cover an interval used to compute Figure~\ref{Fig5}.
\textbf{Right panel}: Bounds on the values of $\alpha$ for negative $n$, showing the
interval of $n$ used calculate Figure \ref{Fig6}.
\label{Fig4} }
\end{center}
\end{figure*}

Now, the standard error deviation
\begin{widetext}
\begin{equation}
\Delta\alpha=\left|\frac{\partial}{\partial H_0}\alpha\right|\Delta
H_0+\left|\frac{\partial}{\partial \rho_0}\alpha\right|\Delta
\rho_0+\left|\frac{\partial}{\partial q_0}\alpha\right|\Delta
q_0+\left|\frac{\partial}{\partial j_0}\alpha\right|\Delta j_0\label{dalpha}
\end{equation}
\end{widetext}
gives rise to lower and upper bounds on the values of $\alpha$ for different values
of $n$. In fact,  making use of the uncertainties of the parameters $H_0$, $q_0$,
$j_0$ and $\rho_0$  (Table~\ref{Table1}) one can plot the curves in the panels
of Figure~\ref{Fig4} to illustrate these bounds%
\footnote{The uncertainties in the speed of light $c$ and Newton constant $G$ are
clearly negligible for our calculations.}.
They have been used to guide suitable choices of values of the
parameters $\alpha$ and $n$ below.

\begin{figure*}[] 
\begin{center}
\includegraphics[width=6.5cm,height=4.0cm]{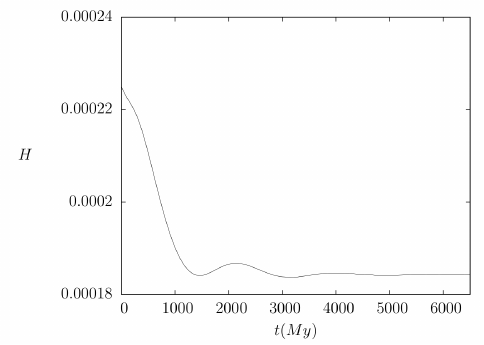}  
\hspace{8mm}
\includegraphics[width=6.5cm,height=4.0cm]{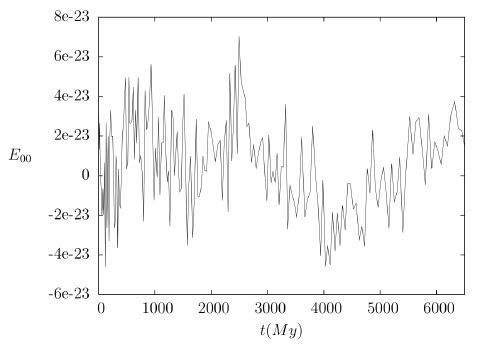}  
\caption{
\textbf{Left panel}: Time evolution of $H$ for spatially flat FLRW dust model in the $f(R)$ theory
defined by Eq.~\eqref{f(R)}. It shows the case in which $n=0.2$ and $\alpha=-4.295\times10^{-6}$.
The initial conditions are the best fit values of the parameters collected together in Table~\ref{Table1}.
\textbf{Right panel}: The behavior of the constraint $E_{00}$ as given by~\eqref{Energia}, which is smaller than $10^{-22}$ for all time $t$, making explicit the precision of the numerical calculation throughout evolution of the model. \label{Fig5}}
\end{center}
\end{figure*}

By applying the same analysis introduced in Section \ref{ABCL-sec} and
using again the main values at Table~\ref{Table1} as initial conditions for
the FLRW dust flat models, we have that for $n>0$, many parameter choices
lead to asymptotic de Sitter solutions, which is  consistent with the results in the
literature~\cite{manyresults}.
As an example, Figure~\ref{Fig5} depicts $H$ for of $n=0.2$ and
$\alpha=-4.295\times10^{-6}$, which are typical values between
the above-mentioned bounds in the left panel of Figure~\ref{Fig4}.
In Figure~\ref{Fig5} we  also show the behavior of the constraint $E_{00}$,
which exhibits fluctuations around zero with rather small amplitudes,
making apparent the degree of confidence  of the numerical calculations.
For this particular value of $\alpha$, $n$ and initial conditions as given by Planck team,
the de Sitter analytical vacuum solution obtained by solving Eq.~\eqref{eq campo}
gives $H=1.8431\times 10^{-4}$, in rather good agreement with the asymptotic values
shown in Figure~\ref{Fig5}. It is possible to show this solution is an attractor over
this phase space region, with $f^{\prime\prime}>0$, as is also the case
for the example depicted before in Figure~\ref{Fig3}.

\begin{figure*}[] 
\begin{center}
\includegraphics[width=6.5cm,height=4.0cm]{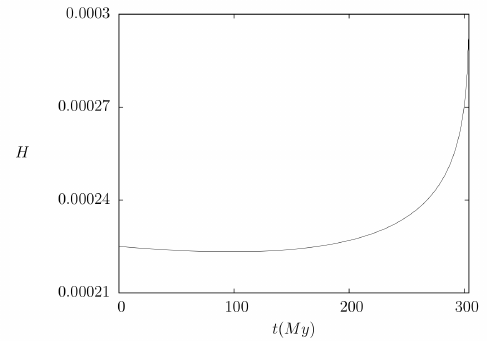}  
\hspace{8mm}
\includegraphics[width=6.5cm,height=4.0cm]{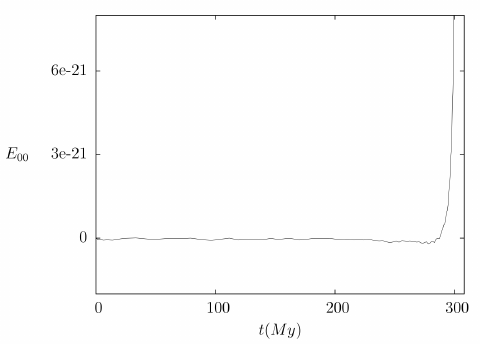}  
\caption{
\textbf{Left panel}: Time evolution of $H$ for spatially flat FLRW dust model in the $f(R)$ theory
given by Eq.~\eqref{f(R)}. It shows the illustrative case in which $n=-0.1$ and $\alpha=-4.557\times10^{-8}$.
The initial conditions were take to be the best fit values collected in Table~\ref{Table1} as given
by Planck team~\cite{CMB}.
\textbf{Right panel}: The behavior of the constraint $E_{00}$ as given by~\eqref{Energia}. As expected,
the constraint increases as the curvature singularity approaches, where the numerical solution must
be halted. \label{Fig6}} 
\end{center}
\end{figure*}

As for the  $n<0$ case, many initial conditions lead to a big rip singularity.
This is a curvature singularity in the sense it cannot be removed by a coordinate
transformation. As an example, we show in Figure~\ref{Fig6} the time evolution of
the Hubble parameter of the dust flat FLRW model for $n=-0.1$ and
$\alpha=-4.557 \times10^{-8}$, chosen by taking into account right panel of
Figure~\ref{Fig4}.
In this case, the analytical vacuum solution to the field equation \eqref{eq campo}
gives  $H=1.8618\times 10^{-4}$. It is also possible to show this solution is a repellor
in this phase space region, rendering the divergent evolution shown in Figure \ref{Fig6}
an expected feature, complemented by our verification that it presents $f^{\prime\prime}<0$
asymptotically.
As indicated by the right panel of Figure~\ref{Fig6}, the numerical constraint $E_{00}$
noticeably increases as the solution approaches the physical singularity, which is
expected from the outset, and then the numerical solution must be truncated.

It is also interesting to compare the future dynamics shown in
Figure~\ref{Fig2}, for ABCL gravity, with that of Figure~\ref{Fig6}, for
$f(R) = R+\alpha R^n$ with $n<0$. The dynamic depicted by Figure~\ref{Fig6}
is a typical example of the run away evolution that suddenly ends in a big rip singularity.
This evolution is usually understood in terms of the ghost-like behavior
due to its transition to a regime where $f^{\prime\prime}<0$.
On the other hand, Figure~\ref{Fig2} presents another runaway solution, that also develops
a ghost-driven regime, but with no sudden singularity within the accuracy of the numerical analysis.
Thus, even when $f^{\prime\prime}<0$ the evolution in one theory (ABCL gravity) presents the remarkable
property of smoothing out the expected divergence. These different behaviors  make apparent
the richness of possible evolutions in the context of $f(R)$ gravity theories.

\begin{figure*}[] 
\begin{center}
\includegraphics[width=6.5cm,height=4.0cm]{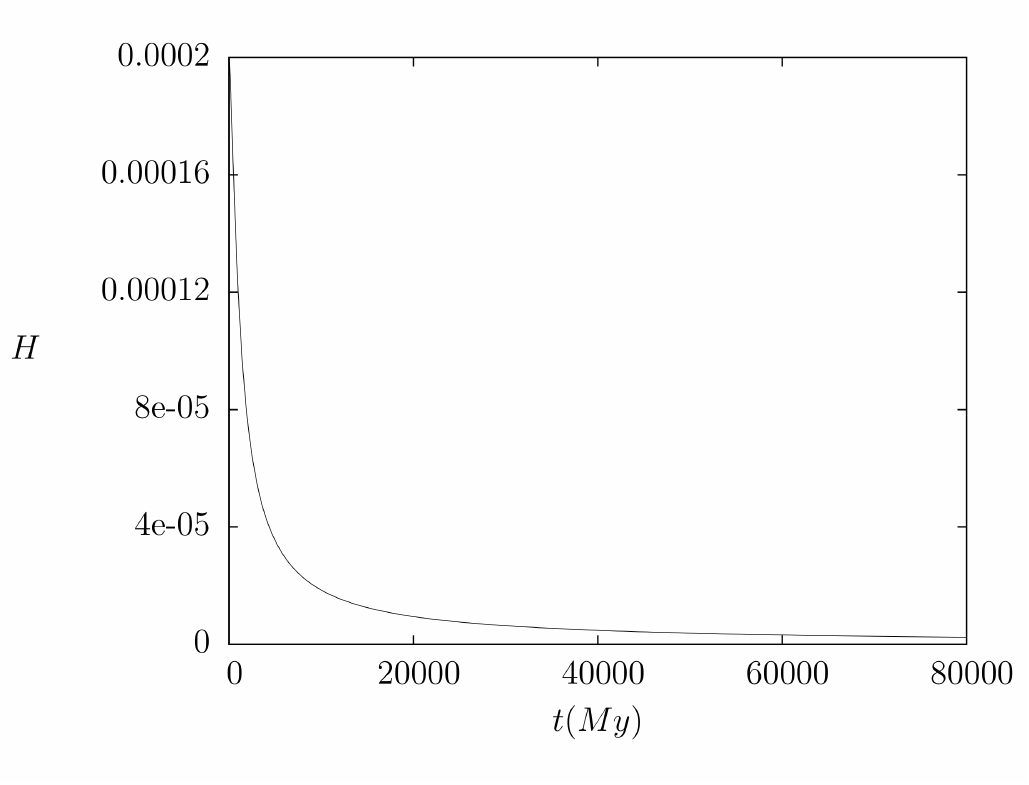}  
\hspace{8mm}
\includegraphics[width=6.5cm,height=4.0cm]{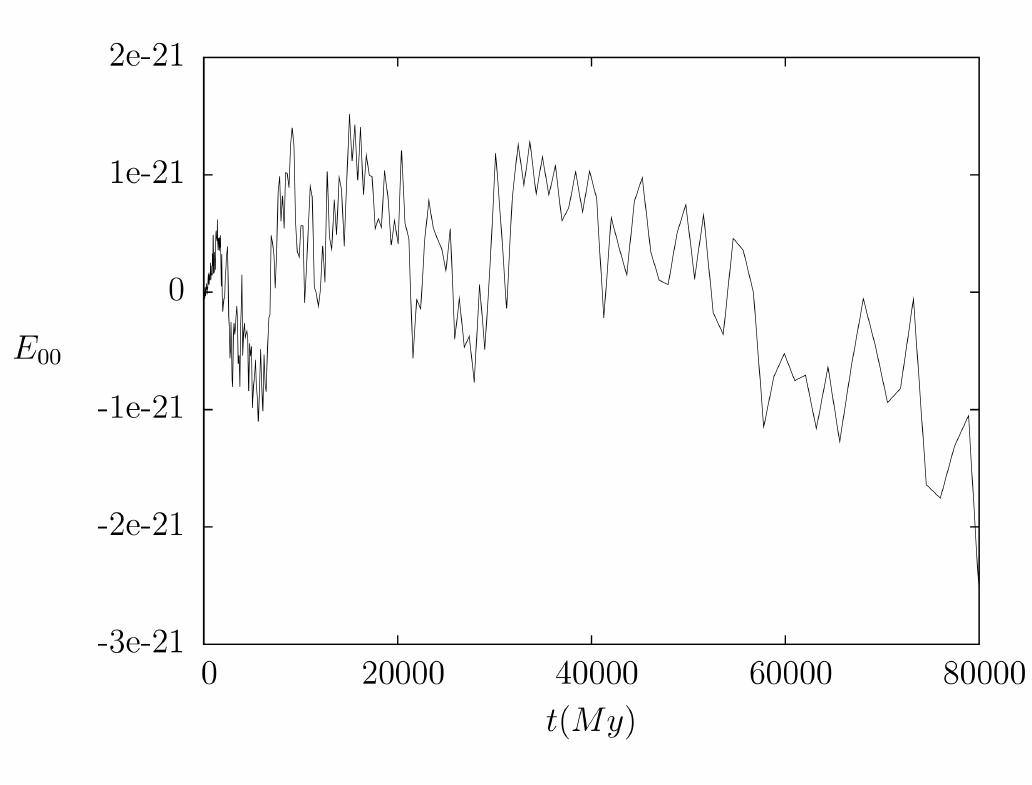}  
\caption{
\textbf{Left panel}: Time evolution of $H$ for spatially flat FLRW dust model in the $f(R)$
theory given by Eq.~\eqref{f(R)}.
As in Figure~\ref{Fig6}, the values of the parameters were taken to be  $n=-0.1$ and $\alpha=-4.557\times10^{-8}$, but now a slightly different value for $H_0$.
Other initial values are given by Table~\ref{Table1}.
It can be seen that Minkowski space is obtained asymptotically when $t\rightarrow \infty$,
as $H\rightarrow 0 = const$.
\textbf{Right panel}: The behavior of the constraint $E_{00}$ as given by~\eqref{Energia},
which clearly is smaller than $10^{-20}$ for all time $t$.
\label{Fig7} } 
\end{center}
\end{figure*}

We can further examine the connection between big rip singularities and ghost-like regimes
($f^{\prime\prime}<0$) through another example. In fact,
by taking the same values for the parameters $n$ and $\alpha$ used for Figure~\ref{Fig6}
along with a slightly different value for the Hubble parameter ($H_0=2.0\times 10^{-4}$,
in units of Table~\ref{Table1}), we have calculated the future dynamics of
the FLRW dust flat models shown in Figure \ref{Fig7}.
For this case, since $\dot{H}\rightarrow 0$ and $H\rightarrow 0$, from Eq.~\eqref{R} one has
the Ricci scalar tends to zero, and thus $f^{\prime\prime}\rightarrow -\infty$.
This evolution illustrates a case where, in a limit strongly associated to ghost-dominated
regimes ($f^{\prime\prime}\ll 0$), the solution actually evolves simply to Minkowski spacetime.
To the best of our knowledge, this interesting dynamical behavior has not been highlighted
so far in the literature.
It also illustrates how a direct association of big rip singularities, or even run away solutions,
with a ghost-like regime of the field equations can be misleading in the framework of
$f(R)$ theories.

\section{Final Remarks and Conclusions \label{conc}}

\begin{figure*}[] 
\begin{center}
\includegraphics[width=9.5cm,height=5.0cm]{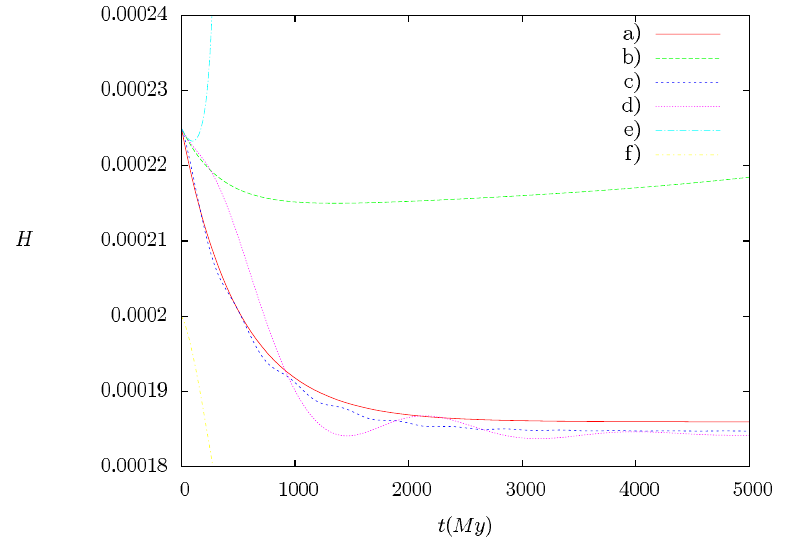}
\caption{Summary of future dynamics numerical analyses for explicit comparison. Except for (f),
they all correspond to the same initial best fit values of the cosmological parameters
collected together in Table~\ref{Table1} as given by Planck team~\cite{CMB}.
$(a)$ Standard FLRW dust flat solution in Einstein's equations, which
tends asymptotically to a de Sitter universe.  
$(b)$  ABCL's theory, eq. \eqref{t_abcl}, when initially
$f^{\prime\prime\prime}\neq 0$, as discussed in subsection \ref{f0neq}. 
$(c)$  A case when initial conditions satisfy $f^{\prime\prime\prime}=0$ in ABCL
theory, as discussed in subsection \ref{f'''=0}. 
$(d)$ Future dynamics of FLRW dust flat model for $f(R) = R+\alpha R^n$ with $n>0$.
$(e)$ Same as $(d)$ but now  with $n<0$.
$(f)$ The same ($\alpha,n$) as in $(e)$, but now with a slightly different value of the Hubble
parameter $H_0$. 
\label{Fig8}}
\end{center}
\end{figure*}

There has been a great deal of recent papers on $f(R)$ gravity motivated by the
attempts to explain the current cosmic acceleration with no need of invoking
a dark energy component.
Despite the arbitrariness in the choice of different functional
forms of $f(R)$, which call for ways of constraining the possible  $f(R)$ gravity theories
on physical grounds, several features of these gravity theories have been discussed
in a number of recent articles.
In this paper we have proceeded further with the investigation of potentialities and limitations
of $f(R)$ gravity theories by examining whether the future dynamics can be used to break
the degeneracy between $f(R)$ gravity theories.
To this end, by taking the recent constraints on the cosmological parameters made
by the Planck team, we have performed a detailed numerical study of the future
dynamic of spatially homogeneous and isotropic dust flat models in the framework of
two gravity theories. 
As a first result, we have shown that  besides being powerful for discriminating between $f(R)$ gravity
theories, the future dynamics numerical technique introduced in this paper
can also be used to determine the fate of the Universe in the framework of these
$f(R)$ gravity theories.
Figure~\ref{Fig8} collects together the results of the future dynamics numerical analyses
of the  FLRW  dust flat models in several different cases. Curve $(a)$ shows the future
evolution of this model in general relativity theory [$f(R)=R)$], while the other curves
represent future dynamics of these FLRW models in several instances as follows.
Curve $(b)$ ABCL theory, Eq.~\eqref{t_abcl}] with $f^{\prime\prime\prime}\neq 0\,$;
Curve $(c)$ ABCL theory with $f^{\prime\prime\prime}=0\,$;
Curve $(d)$ $f(R)= R+\alpha R^n$  with positive $n\,$;
Curve $(e)$ $f(R)= R+\alpha R^n$  with negative $n\,$;
Curve $(f)$ $f(R)= R+\alpha R^n$  with negative $n\,$ but now with a slightly different
value of the Hubble parameter.
Figure~\ref{Fig8} shows that, although with differences for $t\lesssim 1.5 Myr$, the ultimate
fate of the Universe is a de Sitter model (with slightly different Hubble constant) for the
cases $(a)$, $(c)$ and $(d)$. Clearly the cases $e$ evolves to Minkowski space whereas
the case $f$ develops singularity.
Thus, the future dynamics scheme developed in this paper is
indeed a powerful tool to discriminate $f(R)$ gravity theories.

The development of a big rip singularity as shown in curve (e) of Figure~\ref{Fig8}
is consistent with the results found in the literature for $f(R) = R + \alpha R^{n}$
with negative $n$, and are generally associated with ghost-like regimes ($f^{\prime\prime}<0$).
In this regard, an interesting outcome of our analyses is the evolution given by curve
(b) of Figure \ref{Fig8}, where another ghost-driven regime follows a smooth accelerated
evolution with no associated singularity. Along the same lines, we
also note the interesting case presented by curve (f) in the same figure.
This is a solution evolving to Minkowski spacetime for $f^{\prime\prime}$ negative
and unbounded. Thus,  our numerical analyses suggest that ghost-like regimes
($f^{\prime\prime}<0$) do not necessarily lead to singularities.


The discovery of the cosmological expansion along with earlier theoretical
investigations concerning spatially homogeneous and isotropic models by
Friedmann and Lema\^{\i}tre sparked the first scientific studies on the future dynamics
and the ultimate fate of the Universe in the framework of Einstein's theory of gravitation.
It was  shown that if the matter content of the universe is a pressure-free dust, then
the future of the Universe would depend only on the sign of the spatial curvature.
It would expand forever if it has an Euclidean or hyperbolic spatial geometry, and
would expand and eventually recollapse if it has a spherical spatial geometry.
These predictions may be found in virtually any textbook on cosmology, but the
discovery of the accelerating expansion, through type Ia supernovae observations,
made apparent that this simple future forecast for the Universe does not  work
anymore, since the negative-pressure dark energy (DE) component, invoked to account
for the acceleration,  plays a crucial role in the evolution of the Universe.
Indeed, the dark energy is usually described by the equation-of-state
parameter  $\omega$  which is the ratio of the DE pressure to its density
($\omega= p/\rho$). A value $\omega < -1/3$ is required for cosmic acceleration.
When $ -1 < \omega < -1/3$  the DE density decreases with the scale factor $a(t)$.
However, it has been shown that if $\omega <-1$ the dark energy  density becomes 
infinite in a finite-time, $t_s$ (say), driving therefore the universe to a future
finite-time singularity called Big Rip in which when $t \mapsto t_s, \;\,\mbox{then}\;\ \rho \mapsto \infty,
a(t) \mapsto \infty \; \mbox{and} \; |p| \mapsto \infty $~\cite{Caldwell-Starobinsky}.
Afterwards,  it was understood that this is not the only possible doomsday of a dark energy
dominated universe. It may, for example, come to an end in a sudden singularity~\cite{Sudden},
a big freeze doomsday~\cite{Bigfreeze}, or a little rip~\cite{Litterip}.
In this paper we have examined numerically the future dynamics of the Universe but,
instead of assuming a dark energy component, we have assumed that gravity is governed by
$f(R)$ gravity theories and have kept a pressure-free dust as matter content.
As a result, we have also shown  that even if we do not invoke a dark energy
component with $\omega < -1$ one still can have pressure-free dust FLRW flat solution
with a big rip, if gravity deviates from general relativity via
$f(R) = R + \alpha R^n $, as shown in Figure~\ref{Fig6}.

Finally, by using the future dynamics scheme of this paper, we also show an example in
which the ghost-like regimes ($f''<0) $ do not necessarily lead to singularity (Figure~\ref{Fig7}).
Thus, the future dynamics scheme we have developed in this paper is
not only a powerful tool to discriminate between $f(R)$ gravity theories, but
it has also permitted to shed some light on the ghost-like regime and its
connection with singularities in the context of $f(R)$ gravity theories.

\begin{acknowledgments}  
M.J.Rebou\c{c}as acknowledges the support of FAPERJ under a CNE E-26/102.328/2013 grant.
M.J.R. also thanks the CNPq for the grants under which this work was carried out.
D. M\"uller thanks CAPES for the fellowship  Proc.8772-13-4.
\end{acknowledgments}


\end{document}